\documentstyle[11pt,newpasp,twoside,epsf,rotating]{article}
\def\edcomment#1{\iffalse\marginpar{\raggedright\sl#1\/}\else\relax\fi}
\reversemarginpar

\begin{document}
\title{Radio AGN surveys}
\author{Carlos De Breuck}
\affil{Institut d'Astrophysique de Paris, France}
\author{Wil van Breugel}
\affil{IGPP/LLNL, Livermore, CA, USA}
\author{Huub R\"ottgering}
\affil{Sterrewacht Leiden, The Netherlands}
\author{Chris Carilli}
\affil{NRAO, Socorro, NM, USA}

\begin{abstract}
We present a short overview of radio surveys for AGN, including the `complete' flux limited surveys and `filtered' surveys. We also describe our ultra-steep spectrum search for the highest redshift radio galaxies, and our follow-up VLA and ATCA observations of the most distant ($z=5.19$) and the most luminous $z<2$ radio galaxy known.
\end{abstract}

\section{Flux limited and filtered radio surveys}
Radio surveys play a crucial role in the study of AGN. Searches for AGN at radio wavelengths have several major advantages over other wavelength regimes, including: (i) the radio emission is often extremely powerful, and can be detected out to the highest redshifts (e.g. SDSSp~J083643.85+005453.3 at $z=5.82$, Fan et al. 2001), (ii) the most powerful radio sources at the highest redshifts pinpoint the most massive and luminous galaxies at such redshifts, allowing a detailed study, (iii) radio emission is not affected by dust emission, which often affects optical selection criteria.

Several major observational efforts have therefore been performed during the last 4 decades to identify the host galaxies of large samples of radio sources (Table~1). These samples can be divided between `complete' flux limited surveys, and `filtered' surveys, designed to select the highest redshift objects.
The advantage of the flux limited surveys is that, given sufficient spectroscopic redshift information, they can be used to derive the radio luminosity function (RLF; e. g. Dunlop \& Peacock 1990, Willott et al. 2001). This RLF shows a strong increase between $z=0$ and $z \sim 2$, and a possible decrease at larger redshifts is subject of considerable debate (e.g. Jarvis et al. 2001b, Waddington et al. 2001).
The main problem for establishing the high-z RLF is the small number of $z>2$ radio galaxies and quasars in the flux limited samples.  

To find more high redshift radio galaxies (HzRGs; $z>2$), additional selection criteria have been used to construct `filtered' surveys. The most successful of these is the selection of sources with ultra steep radio spectra (USS), although others such as angular size upper limits have also been used. While the main explanation for the success of the USS criterion is a simple $k-$correction of the generally concave radio spectrum, other effects could strengthen the $\alpha - z$ correlation, including (i) the steepening of the rest-frame radio spectral index with radio power, and (ii) more important inverse Compton losses at high redshift.

Table~1 shows that these filtered surveys reach fainter flux densities, and find objects with a mean redshift $z \sim 2$, while the flux limited surveys target brighter sources at $z \sim 1$.

\section{An ultra-steep spectrum search for high redshift radio galaxies}
With the advent of several major radio surveys during the last decade (Table~2), it is now possible to construct large, uniform samples of radio sources. We have started a program to find significant numbers of $z>3$ radio galaxies. Our 3 samples consist of 669 USS sources, covering the entire sky outside the Galactic plane (see De~Breuck et al. 2000). To derive spectral indices, we combined the WENSS and TEXAS survey with the NVSS. Figure~1 and Table~2 show that WENSS is ideally matched in both resolution and sensitivity for such a USS sample. In the regions not covered by the WENSS survey ($\delta < +28$\deg), we have used the shallower TEXAS survey. 

\begin{figure}
\plotfiddle{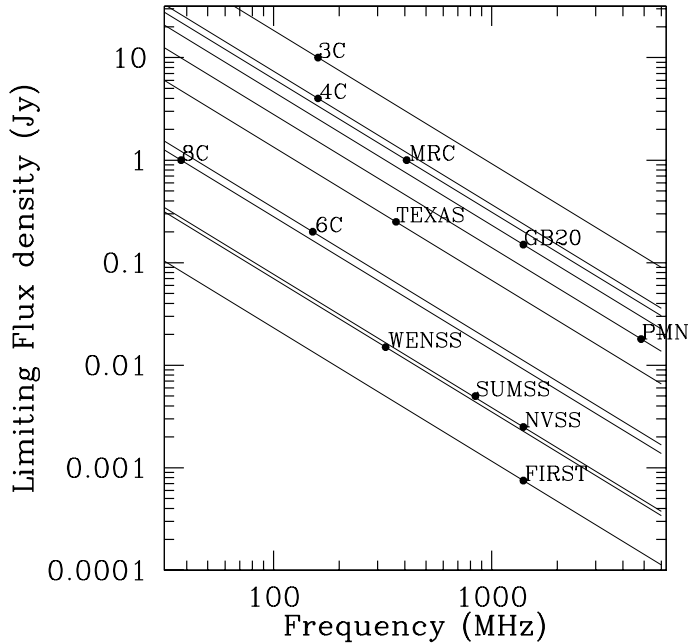}{7.8cm}{0}{45}{45}{-140}{-70}
\caption{
Limiting flux density of all major radio surveys. Lines are of constant spectral indices of $-1.3$. Note that WENSS and SUMSS are ideally matched to NVSS to construct samples of USS sources.}
\end{figure}

More recently, we have also constructed a deeper sample in the $-25\deg<\delta<-8\deg$ region using the southern extension of the WENSS, the Westerbork in the Southern Hemisphere (WISH, De~Breuck et al., in prep.). At $-40\deg<\delta<-30\deg$, we shall combine the SUMSS with the NVSS to construct a first sensitive USS sample in the deep southern hemisphere.

At $\delta < -40\deg$, the extragalactic radio sky is even less explored, with several extremely luminous radio sources (comparable to northern 3C sources) remaining to be identified. Burgess \& Hunstead (1994) describe the only observational effort to date to identify these intriguing objects.
We have therefore also constructed the first USS sample at $\delta < -40\deg$, using the MRC and PMN surveys. From this sample, we have already discovered one of the most powerful radio sources known: MP~J2045$-$6018 at $z=1.464$ (Fig.~2). 

\begin{figure}
\plotfiddle{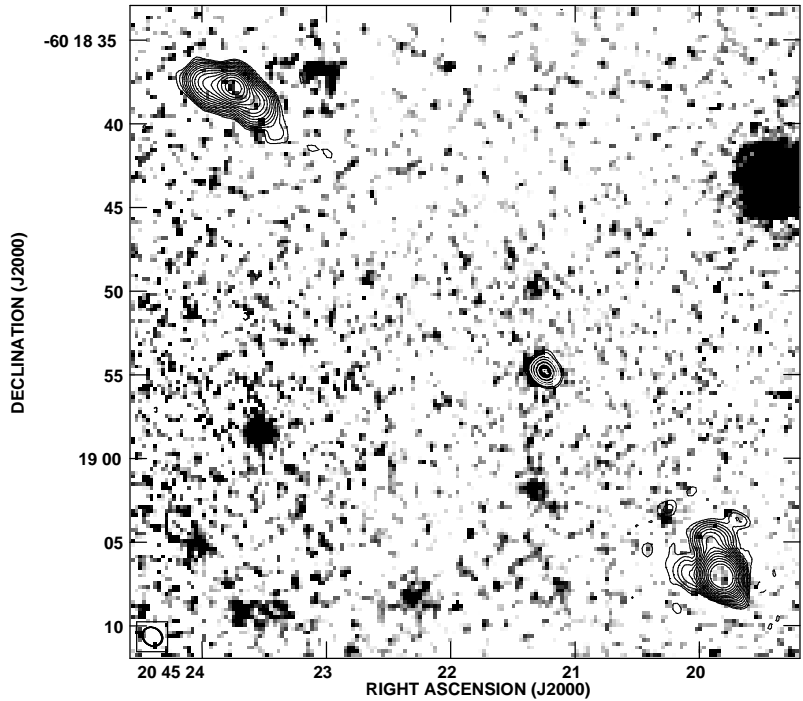}{9cm}{270}{50}{50}{-200}{290}
\caption{
MP~J2045$-$6018 ($z=1.464$): {\it Greyscales:} CTIO $K-$band image; {\it Contours:} deep ATCA 8.64~GHz image. This is the most powerful radio source known at $z<2$. Note the asymmetric structures in the radio lobes, and the weak radio core. The separation between the radio lobes is 32\farcs6.}
\end{figure}

To identify the radio sources in our USS samples, we first obtained high resolution radio images from the VLA and ATCA. After an initial campaign of optical identifications, we switched to $K-$band imaging using the Keck and CTIO telescopes (De~Breuck et al. 2002). Even with moderately deep imaging ($K<$20 to $K<$22), we find an identification rate of $>$95\% for the 86 sources observed in $K-$band. Subsequent optical spectroscopy of 46 sources with 3-10m class telescopes has yielded redshifts for 72\% of these, with a mean redshift of $z \sim 2.5$ (De~Breuck et al. 2001).
From this sample, we have discovered the most distant radio galaxy known: TN~J0924$-$2201 at $z=5.19$ (van Breugel et al. 1999). Figure~3 shows the radio spectrum determined from multi-frequency VLA imaging. The spectrum curves at high frequencies, consistent with what is seen in most powerful radio galaxies. Our deep VLA images (Fig.~3) show a 1\farcs25 double radio source, but no radio core. 

\begin{figure}
\plotone{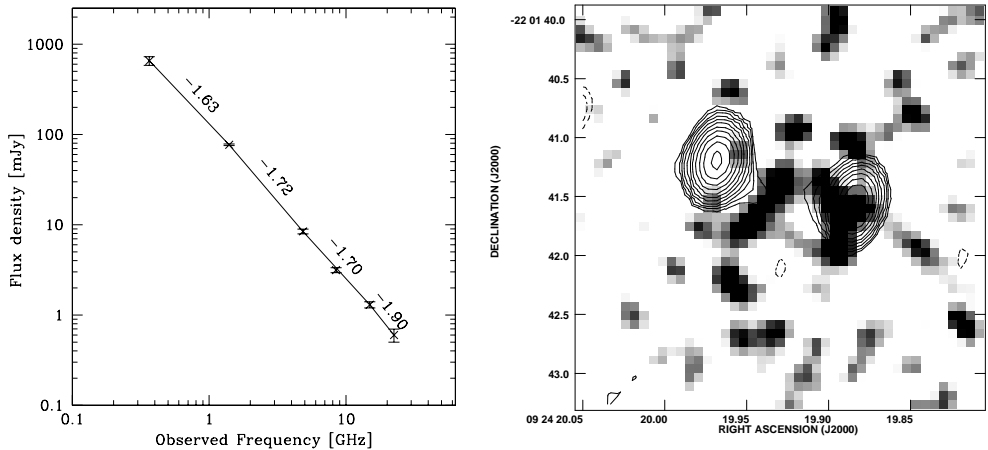}
\caption{
TN~J0924$-$2201 ($z=5.19$): {\it Left:} radio spectrum, based on VLA observations. Note the steeper spectral indices at higher frequencies. {\it Right:} VLA 8.6~GHz map overlaid on a Keck/NIRC $K-$band image. The relative astrometric uncertainty is $<0\farcs2$.}
\end{figure}

Interestingly, despite integration times of 1h or more with the Keck telescope, 5 sources show only a faint continuum emission, but no identifiable emission lines between 4000~\AA\ and 9000~\AA. These sources are possibly obscured AGN, as suggested by the detection of several of these at (sub)mm wavelengths (Reuland et al., in preparation). 

\acknowledgements
This work was supported by a Marie Curie Fellowship of the European Community programme `Improving Human Research Potential and the Socio-Economic Knowledge Base' under contract number HPMF-CT-2000-00721, and by the Research and Training Network `The Physics of the IGM' set up by the Human Potential Programme of the European Commission.
The work by WvB was performed under the auspices of the U.S. Department of Energy, National Nuclear Security Administration by the University of California, Lawrence Livermore National Laboratory under contract No. W-7405-Eng-48.

\begin{figure}
\plotfiddle{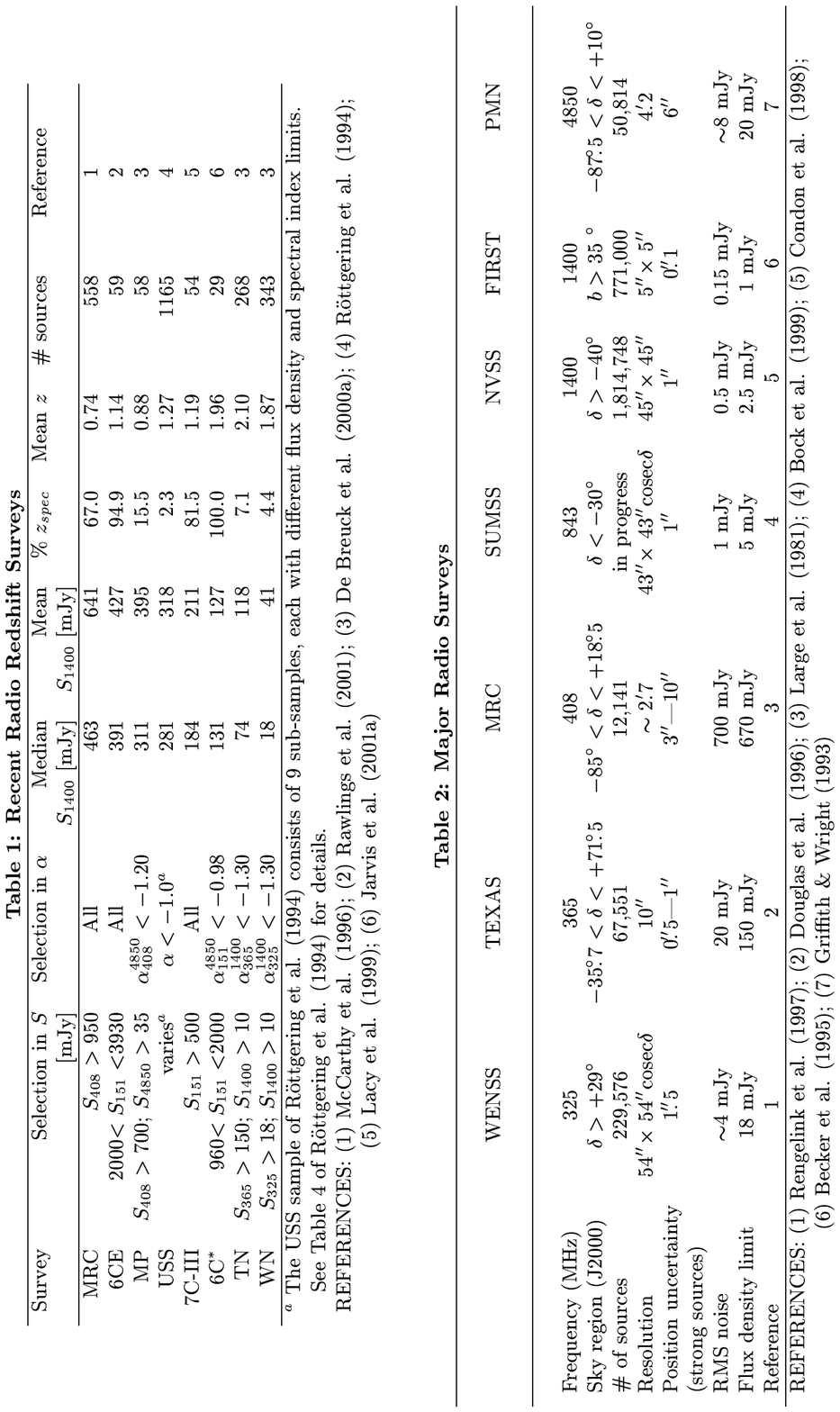}{15cm}{0}{88}{88}{-250}{-70}
\end{figure}

\end{document}